\documentclass[polish,openany,plhyphen,11pt]{article}
\usepackage{amssymb}
\date{ }
\title{What was the temperature of the Bagsik financial
oscillator?
}
\author{E. W. Piotrowski\\ Institute of Theoretical Physics,
University of Bia\l ystok,\\ Lipowa 41, Pl 15424 Bia\l ystok,
Poland\\ e-mail: ep@alpha.uwb.edu.pl\\ and J. S\l adkowski
\\ Institute of Physics, University of Silesia, \\ Uniwersytecka
4, Pl 40007 Katowice, Poland \\ e-mail: sladk@us.edu.pl}
\begin{document}
\maketitle
\def\Z{{\bf Z\!\!Z}}
\def\R{{\bf I\!R}}
\def\N{{\bf I\!N}}
\begin{abstract}
We argue that the recently published by Przystawa
and Wolf model of the Bagsik financial oscillator is
oversimplified and unrealistic. We propose and analyze a refined
explanation of this rare financial phenomenon. We have found an
example that results in profitability about 45 000 times bigger
than that of the Przystawa and Wolf model.

\end{abstract}

 \vspace{5mm}
\section{Introduction} In a recent paper \cite{1} Przystawa and
Wolf discuss an algorithm (denoted below as $sbO$) that, if
exploited in the plunged in hyper-inflation  Poland of the early
nineties, should bring enormous profits. This algorithm  make the
most of constancy of the exchange ratio between  two currencies,
dollar (USD) and Polish z\l oty (PLN), and interest rates.
Przystawa and Wolf suggest that Polish contractor Bagsik made his
enormous fortune by exploiting such mechanisms (a version of
financial oscillator denoted $sbO_{0.7}$, see below). We would
like to argue that this is not justified because the real profits
from the use of the oscillator $sbO_{0.7}$ do not result in
substantial magnification of the capital. We propose a different
explanation of the Bagsik rapid enrichment based on a different
capital source (the appropriate oscillator will be denoted as
$sbO_{1.93}$). Due to the polemic character of the article we will
refer to numerical data and time spread discussed in Ref. [1].
\section{Logarithmic discount rates} To determine discount rates
(or interest rates for deposits) in the interval between the
moments $k,m\in\mathbb{N}$, $k<m$  banks use a discount factor
$U(k,m)\in\mathbb{R}_+$. The lengths of the intervals in question
are not necessary the same if measured in physical units of time.
This means that  the bank lends the amount of 1 at the moment $k$
on return of the amount of $U_c(k,m)$ at the moment $m$.
Analogously, if the amount of 1 is deposited at the moment $k$
then the bank gives back the amount of $U_d(k,m)$  at the moment
$m$. Profitability of bank activities implies the inequality
$U_c(k,m)>U_d(k,m)$. The discount factor is a monotone function,
$U(m,m+k)>1$, and fulfills the condition of multiplicativity
$U(k,l)U(l,m)=U(k,m)$. Of course, $U(m,m)=1$. It is convenient to
make use of the notion of logarithmic rates $R(k,m):=\ln U(k,m)$
because their properties are more legible and calculations are
simpler. The appropriate properties take the form $R(m,m+k)>0$,
$R(m,m)=0$, and
\begin{equation}
\label{addytyw} R(k,l)+R(l,m)=R(k,m) .
\end{equation}

\section{The slow bond oscillator} Let us suppose that an
arbitrageur has at his disposal two banks, $A$ and $B$. The first
one is ready to lend on the basis of the logarithmic rate
$R_A(k,m)$. The second one accepts deposits on the basis of the
rate $R_B(k,b)$. In addition let $ R_A(m,m+k)\ll R_B(m,m+k)$. The
arbitrageur aims at borrowing capital from $A$ and depositing the
capital in $B$ so that the financial gain will be the highest
possible. The authors of Ref. [1] focused their attention on the
following algorithm ($sbO$) which, in  their opinion, should
explain Bagsik unheard-of financial achievements in Poland during
1990.
\begin{description}
\item[ moment $0$:] {\it
The banker $A$ estimates that the assets of $X$ (say his premises)
would be worth 1 at the moment $N$, so he lends him  the amount of
$e^{-R_A(0,N)}$ (say a mortgage loan). The arbitrageur $X$ is
obliged to give back $A$ the amount of 1 at the moment $N$. The
banker $B$ offers for a deposit of 1 at the moment 0 the amount of
$e^{R_B(0,N)}$ to be paid at the moment $N$. This means that the
banker $B$ enters into the obligation to pay $X$
\begin{equation}
\label{poczat} p_0:=e^{R_B(0,N)-R_A(0,N)}\vspace{-4ex}
\end{equation}
 \\
at the moment $N$} which is testified by issuing an appropriate
bond to $X$.
\end{description}

\begin{description}
\item[ moment $k$:] {\it
By accepting the new bond, the banker $A$ finds out that the
present revealed assets of $X$ (he already has bonds for
previously revealed assets) will be worth $p_{k-1}$ at the moment
$N$. Therefore $A$ pays the amount of $e^{-R_A(k,N)}p_{k-1}$ to
$X$. The banker $B$, following the above rules, accepts the
deposit of $e^{-R_A(k,N)}p_{k-1}$ and issues the bond to return
the additional amount of

\begin{equation}
\label{kwoty} p_k:=e^{R_B(k,N)-R_A(k,N)}p_{k-1}\vspace{1ex}
\end{equation}

to $X$ at the moment $N$. }
\end{description}
The multiple issued by the banker $B$ bond (certificate) allows
the arbitrageur $X$ to retrieve the stated amount of money from
$B$. The recurrence formula $(\ref{kwoty})$ states the banal fact
that to know the figures stated on the $k$-th bond it is
sufficient to multiply the amount from the previous one by the
capitalization factor $e^{R_B(k,N)-R_A(k,N)}$. The initial
condition $(\ref{poczat})$ leads to
$p_{N-1}=e^{\sum_{m=0}^{N-1}\bigl(R_B(m,N)-R_A(m,N)\bigr)}$ stated
on the last, issued just before end the arbitrage, bond. This is
the only one bond that is not forwarded to $A$. The rest of the
issued by $B$ bonds is used for securing  the obligations of $X$
with respect to $A$ originated at the moments
$k=1,\ldots,N\negthinspace-\negthinspace1$. If $X$ buys a mortgage
pledge from $A$ at the moment $N$ then he has, besides the
premises, the funds
\begin{equation}
\label{wynik} e^{\sum_{m=0}^{N-1}\bigl(R_B(m,N)-R_A(m,N)\bigr)}-1
\end{equation}
at his disposal. It is worth noticing that, the banks $A$ and $B$
may be physically different market institutions. The whole
property of $X$ is worth
$e^{\sum_{m=0}^{N-1}\bigl(R_B(m,N)-R_A(m,N)\bigr)}$ at the moment
$N$ so the logarithmic rate of return of the arbitrage is
\begin{equation}
\label{formu}
\mathfrak{R}_{BA}(0,N)=\sum_{m=0}^{N-1}\Bigl(R_B(m,N)-R_A(m,N)\Bigr).
\end{equation}
We set it in  a different type to denote that $\mathfrak{R}_{BA}$
are not additive. The lack of additivity characterizes all
aggressive techniques of arbitrage. The additivity of the rates
$R(k,m)$ allows to simplify the formula $(\ref{formu})$

\begin{eqnarray}
\label{formu1}
\mathfrak{R}_{BA}(0,N)&=&\sum_{m=0}^{N-1}\sum_{k=m}^{N-1}\Bigl(R_B(k,k+1)
-R_A(k,k+1)\Bigr)\\
&=&\sum_{k=1}^Nk\Bigl( R_B(k-1,k)-R_A(k-1,k)\Bigr).
\end{eqnarray}
If all the intervals are uniformly distributed, that is  the
differences $R_B(k-1,k)-R_A(k-1,k)$ are equal then
$R_B(k-1,k)-R_A(k-1,k)=\frac{1}{N}\Bigl(R_B(0,N)-R_A(0,N)\Bigr)$
and the appropriate rate $\overline{\mathfrak{R}}_{BA}(0,N)$  is
given by
\begin{equation}
\label{wzorek}
\overline{\mathfrak{R}}_{BA}(0,N)=\frac{N+1}{2}\Bigl(R_B(0,N)-R_A(0,N)\Bigr).
\end{equation}
We will call such oscillators uniform $sbO$s. The assets of $X$
who accomplishes a uniform $sbO$  are given by the formula (cf.
$(\ref{wynik})$)
\begin{equation}
e^{\frac{N+1}{2}(R_B(0,N)-R_A(0,N))} -1
\end{equation} which is equivalent to the one given in Ref. [1] (Eq.
(16)). If $R_B(0,N)$ is the highest available in discussed
interval deposit rate then the uniform arbitrage is profitable
under the condition that $\overline{\mathfrak{R}}_{BA}(0,N)$ is
greater than $R_B(0,N)$ which implies
$N>\frac{R_B(0,N)+R_A(0,N)}{R_B(0,N)-R_A(0,N)}$. Let us note  that
the profit given by $(\ref{formu})$ may be achieved only if there
is a closing warranty ($CW$), that is a possibility of
instantaneous transfer of all bonds and the related capitals at
the moment $N$.

\section{Bagsik oscillator}
A physicist would probably say that the presented method of
arbitrage ($sbO$) resembles less  an oscillator than the repeating
mechanism of a mysterious heat engine driven by two thermal baths
with the temperatures $R_B$ and $R_A$, respectively. The highest
efficiency of such engines (without changing constructions) is
reached for $R_B=R_{\min}$ and $R_A=R_{\max}$ that is for the,
respectively, lowest and highest rates of return during a given
interval. Przystawa and Wolf claim that the mechanism
$sbO_{0.8-0.1}$, with the thermal bath in the shape of a deposit
in a Polish bank (the currency PLN, the one year rate $R_B=0.8$)
and the reservoir created by a credit (the currency USD and the
one year rate $R_B=0.1$). The exchange ratio of PLN to USD was
constant during that time. The two currencies were needed only to
show that the accomplishing of the oscillator was impossible
without the constancy of exchange ratio. They forgot that there
were available more interesting "financial thermostats" at that
moment. The present authors think that the hottest $R_{\max}$ and
the coldest $R_{\min}$ rates were offered by the hyper-inflation
itself. The average prices of non-edible goods raised by 591.2\%
according to the official state data \cite{2} which gives the
logarithmic rate $R_{\ast}$ equal to $\ln 6.912\simeq1.93$ (with
respect to PLN). The prices of services raised much more: by
780.7\% which gives $R_{\max} =\ln 8.807\simeq2.18$. The rate
$R_A=R_{\min}=0$ was also available: it was possible  not to repay
an interest free debt in PLN because the undergoing revolutionary
changes Polish law did not offer any mechanism of execution of
debts revalued by the inflation rate at that moment. Let us select
the prices of non-edible goods as the "heat source" of the
oscillator $R_b=R_{\ast}$ (it seems to be difficult to use
services for doing this). The so defined oscillator $sbO_{1.93}$
had a closing warranty build-in. The bank $A$ formed sellers and
the role of the bank $B$ was performed by a belonging to $X$ firm.
$X$ simply put off the due payment for the purchased goods till
the moment $N$. The owned by $X$  firm formed a reservoir of goods
and immovables  any other activity (e. g. production) was
inessential. At the moment $N$ the execution of $CW$ was
immediate: one queue formed horrified creditors and a second one
formed consumers wanting to get rid of theirs cash. The circle was
closed by the lawful deferred payment (one could induce directors
of state-owned firms to enter such formally legal but tragic in
effects contracts). The generally accessible archive of the Polish
internet journal {\it Donosy} \cite{3} reports that at the
beginning of the year 1990 (2 of January) the interests of demand
deposits were at the level of 7\% a year and the three-tears
deposits -- 38\%. Only at the end of the year (13 of December) the
interest rates of the one-year deposits raised to 60\%. Therefore
if we take that a PLN deposit gave a return of 50\% on average in
1990 the number would be overestimated. The logarithmic rate of
such deposits was not $R_B(0,N)=0.8$, as is supposed in the Ref.
[1], but $\ln 1.5\simeq0.4$. This means that the suggested
mechanism led to  Bagsik's return described by the oscillator
$sbO_{0.3}$ and not by $sbO_{0.7}$. The profits given by the
formula ($\ref{wynik}$) for the oscillators $sbO_{0.7}$,
$sbO_{0.3}$ and $sbO_{1.93}$ are presented in the Table 1. The
first column is also given in the Ref. [1]. Note that for $N=12$
the return of $sbO_{1.93}$ is about 45 000 times bigger than that
of $sbO_{0.3}$ !

\begin{center}
 \begin{table}[h]
 \label{tab1}
 \begin{tabular}{rr@{.}lr@{.}lr@{.}l} \hline
  % after \\ : \hline or \cline{col1-col2} \cline{col3-col4} ...
$\phantom{.}_{
R_B(0,N)-R_A(0,N)=}$&\multicolumn{2}{c}{\vphantom{$\Bigl(\frac{A}{A}1$}
\phantom{aga}$0.7$\phantom{a}}&
\multicolumn{2}{c}{\phantom{a}$0.3$}&
\multicolumn{2}{c}{\hspace{-2em}$1.93$}\\
\hline
  \vphantom{\Huge A}$N=$~~1\phantom{agata}& \phantom{agata}1&0138&
     \phantom{agata}0&34986&\phantom{aga}5&8895\phantom{agata}  \\
  2\phantom{agata}& 1&8577&\phantom{agata}0&56831&\phantom{aga} 17&084 \\
  3\phantom{agata}& 3&0552&\phantom{agata}0&82212&\phantom{aga}46&465 \\
   4\phantom{agata}& 4&7546 &\phantom{agata}1&1170&\phantom{aga}123&59 \\
    5\phantom{agata}& 7&1662 &\phantom{agata}1&4596&\phantom{aga} 326&01 \\
     6\phantom{agata}& 10&588 &\phantom{agata}1&8577&\phantom{aga} 857&34\\
      7\phantom{agata}& 15&445 &\phantom{agata}2&3201&\phantom{aga}2252&0 \\
       8\phantom{agata}& 22&336 &\phantom{agata}2&8574&\phantom{aga} 5912&5 \\
        9\phantom{agata}& 32&116 &\phantom{agata}3&4817&\multicolumn{2}{c}{\hspace{-3.14em}15521}  \\
        10\phantom{agata}&45&993 &\phantom{agata}4&2070& \multicolumn{2}{c}{\hspace{-3.14em}40740}\\
          11\phantom{agata}&65&686
          &\phantom{agata}5&0497&1&0694$\times10^5$\\
           12\phantom{agata}&93&632 &\phantom{agata}6&0287&2&8069$\times10^5$
         \vspace{-2ex}\\
  \\ \hline
\end{tabular}
\caption{Profits made from a unit of capital for the arbitrage
$sbO$}
 \end{table}
\end{center}

\section{The slow cash oscillator} An arbitrageur performing
$sbO$ wastes a substantial amount of time between the moments
$k\negthinspace-\negthinspace1$ and $k$
($k=1,\ldots,N\negthinspace-\negthinspace1$) on delivering bonds
to the banker $A$. We will denote the average amount of time
needed for this delivery by $\tau_{B\rightarrow A}$. The authors
of \cite{1} suggest the possibility of realization of an arbitrage
$sbO$ in Poland of 1990 if the bank $A$ gives credits in USD and
the bank $B$ accepts deposits in PLN. Only the bank $B$ could have
operated on the territory of Poland because credits in USD where
then unavailable. Therefore, for obvious reasons, the interval
$\tau_{B\rightarrow A}$ was considerably shorter than
$\tau_{A\rightarrow B}$ during which the arbitrager $X$
transports, avoiding interference from the more and more
suspicious customers, more and more capital from $A$ to $B$ in the
shape of goods or cash. (The interval $\tau_{A\rightarrow B}$ is
equal zero for the discussed in the previous section oscillator
$sbO_{1.93}$ because $B=X$.) We will ignore the necessity of
showing the source of $CW$ and suppose that it was known to the
authors of the Ref. [1]. If $\tau_{A\rightarrow
B}\gg\tau_{B\rightarrow A}\simeq0$ then the algorithm $sbO$ should
be replaced by the following one ($scO$):
\begin{description}
\item[moment $0$:]{\it The banker $A$
estimates that the assets of $X$  would be worth 1 at the moment
$N$, so he lends him  the amount of $p_0=e^{-R_A(0,N)}$.} The
arbitrageur $X$ is obliged to give back $A$ the amount of 1 at
the moment $N$.
\item[moment $k$:]{\it The
banker $B$ offers for a deposit of $p_{k-1}$ at the moment $k$ the
amount of $e^{R_B(k,N)}p_{k-1}$ in the form of a bond becoming due
at the moment $N$. If $k<N\negthinspace-\negthinspace1$ then the
banker $A$  takes this bond as a deposit and pays to $X$ the
amount $p_k:=e^{R_B(k,N)-R_A(k,N)}p_{k-1}$. }
\end{description}
If we take for granted the existence of $CW$ then by repeating the
calculation performed for $sbO$ we easily get the profit made by
$X$ from the arbitrage
\begin{eqnarray*}
&\ &e^{R_B(N-1,N)}\prod_{k=1}^{N-2}e^{R_B(k,N)-R_A(k,N)}e^{-R_A(0,N)}-1\\
&=&e^{\sum_{k=0}^{N-1}(R_B(k,N)-R_A(k,N))}e^{R_A(N-1,N)-R_B(0,N)}-1.
\end{eqnarray*}
The profit is smaller than $\mathfrak{R}_{BA}(0,N)$ because it
equals
\begin{equation}
\label{nowa} \mathfrak{R}_{BA}(0,N)-(R_B(0,N)-R_A(N-1,N)).
\end{equation}
In the case of a uniform arbitrage the logarithmic rate of return
is $\frac{N+1}{2}(R_B(0,N)-R_A(0,N))-R_B(0,N)+\frac{R_A(,N)}{N}$.
Therefore the hypothetical two-currencies variant of the Bagsik
oscillator with $CW$ should result in smaller profits than those
of the oscillator $sbO_{0.3}$ (presented in the Table 1). The
non-multiplicative capitalization coefficient,
$\mathfrak{U}(0,N)$, for $scO_{0.3}$ ($N\leq12$) is smaller
26-36\% than the one corresponding to the arbitrage $sbO_{0.3}$
($e^{R_A(0,1)-R_B(0,1)}\simeq0.741$,
$e^{R_A(11,12)-R_B(0,12)}\simeq0.676$). Effectively, the profit is
the same as in $sbO_{0.3}$ but shortened by one step. \\We may
consider a whole one-parameter family of arbitrage procedures
$\lambda O$ for $N$ full cycles, where $\lambda\in[0,1]$ is the
quotient of lengths of the characteristic intervals, $\lambda =
\frac{\tau_{A\rightarrow B}}{\tau_{B\rightarrow A}}$,
$\tau_{A\rightarrow B}+\tau_{B\rightarrow A}$=1. For example, in
the case when $X$ obtains from $A$ a letter of credit ( a document
issued by $A$ authorizing the bearer to draw money  from another
bank at once) then the intervals $\tau_{A\rightarrow B}$ i
$\tau_{B\rightarrow A} $may be equal (the arbitrage
$\frac{1}{2}O$). We have already discussed two representatives of
the family $\lambda O$ because $0O\negthinspace =\negthinspace
sbO$ and $1O\negthinspace = \negthinspace scO$. Note that the
logarithmic rate of return is a decreasing function of $\lambda$.
This follows from the fact that the greater the $\lambda$ is, the
smaller is the length of the whole time of using the heat
reservoir $B$. So the algorithms $sbO$ and $scO$ give the extreme
values of profits possible by carrying out one of the procedures
$\lambda O$.
\section{The temperature of an arbitrage} The present authors have
proposed to use temperature, that is the Lagrange multiplier
$T^{-1}$  as a measure of the financial gain \cite{4}. This
parameter allows to compare financial achievements on different
market and during different time scales. The thermodynamically
conjugated to the temperature entropy allows to measure qualities
of a financial expert or adviser. The market analyzed in this
paper corresponds to a two level physical system with the energies
$-R_{\max}$ end $-R_{\min}$. We assign, according to the maximal
entropy principle, to groups of investors achieving equal
logarithmic rates of return $R$ a representative canonical
ensemble. The temperature $T^{-1}$ of the ensemble is given by the
following function of $R$ \cite{4}
\begin{equation}
\label{temper1} T^{-1}_{R}= \ln\frac{R-R_{\min}}{R_{\max}-R},
\end{equation}
where $R_{\min}$ and $R_{\max}$ are the lowest and the highest
rates in the considered interval, respectively. We may use the
formula $(\ref{temper1})$ for all real values of $R$ after fixing
the branch of the logarithm. For the rates
$\mathfrak{R}\notin[R_{\min},R_{\max}]$ we get
\begin{equation}
T^{-1}_{\mathfrak{R}}=
\ln\Bigl((-1)\cdot\frac{\mathfrak{R}-R_{\min}}{\mathfrak{R}-R_{\max}}\Bigr)=
i\pi +  \ln\frac{\mathfrak{R}-R_{\min}}{\mathfrak{R}-R_{\max}} .
\end{equation}
Note that contrary to the additive rates case \cite{4} the
presently discussed arbitrage should be prized the more the lower
the real part of the temperature $T^{-1}$ is. May by we should
call financial oscillators only those arbitrages with non-zero
imaginary parts of the temperature? If we determine the proposed
temperatures for the oscillators $sbO_{0.3}$ (the second column of
the Table 1) and $sbO_{1.93}$ (the third column of the Table 1)
then we get the results presented in the Table 2.

\begin{center}
 \begin{table}[h]
 \label{tab2}
 \begin{center}
 \begin{tabular}{rr@{.}lr@{.}l} \hline
  % after \\ : \hline or \cline{col1-col2} \cline{col3-col4} ...
$\phantom{.}_{
R_B(0,N)-R_A(0,N)=}$&\multicolumn{2}{c}{\vphantom{$\Bigl(\frac{A}{A}1$}
\phantom{aga}$0.3$\phantom{a}}&
\multicolumn{2}{c}{\phantom{a}$1.93$}\\
\hline
  \vphantom{\Huge A}$N=$~~1\phantom{agata}& \phantom{agata}-1&8352&
 \multicolumn{2}{c}{\hspace{2.7em}2.0438$=T_{\ast}^{-1}$}  \\
  2\phantom{agata}& -1&3466 &\phantom{agata} $i\pi +$1&3985 \\
  3\phantom{agata}& -0&96825 &\phantom{agata} $i\pi +$0&83187  \\
   4\phantom{agata}& -0&64536 &\phantom{agata} $i\pi +$0&60114 \\
    5\phantom{agata}& -0&35222 &\phantom{agata} $i\pi +$0&47243 \\
     6\phantom{agata}& -0&073428 &\phantom{agata} $i\pi +$0&38968 \\
      7\phantom{agata}& 0&20252 &\phantom{agata} $i\pi +$0&33182 \\
       8\phantom{agata}& 0&48643  &\phantom{agata}$i\pi +$0&28903 \\
        9\phantom{agata}& 0&79113 &\phantom{agata}$i\pi +$0&25606   \\
        10\phantom{agata}& 1&13565 &\phantom{agata}$i\pi +$0&22988 \\
          11\phantom{agata}& 1&5554 &\phantom{agata}$i\pi +$0&20857 \\
           12\phantom{agata}& 2&1375 &\phantom{agata}$i\pi +$0&19089 \vspace{-2ex} \\
  \\ \hline
\end{tabular}
\end{center}
\caption{Temperatures $T^{-1}$ of the  $sbO$ arbitages}
 \end{table}
\end{center}
Negative temperatures characterize financial activities
unprofitable  even on a developed efficient market \cite{4}. It is
worth to note that the temperatures $T^{-1}$ lower than
$T^{-1}_{\ast}\negthinspace=\negthinspace2.0438$ (see the Table 2)
also correspond to disadvantageous achievements because during
that period the temperature $T^{-1}_{\ast}$ was easily achieved by
every citizen of Poland who possessed goods of everyday use (and
no local money). Therefore the first column of the Table 1
presents doubtful financial achievements. For $scO_{0.3}$ with
$CW$ the profit is positive only if
$N\negthinspace=\negthinspace13$ what call in question the
possibility of using this oscillator as a tool in making capital
in Poland of the early nineties. And we have neglected the
substantial  starting and clearing costs of such an arbitrage! It
would be interesting to know if and to what extent arbitrages of
the type $sbO_{0.3}$ implemented by Polish banks served as a
driving force of the hyper-inflation. Such an oscillator might
consist in giving credits in Polish z\l oty  and accepting
deposits in foreign bills. The inflation was brought under control
simultaneously with the exhaustion of foreign currencies savings
of the population. It seems that this substantially slackened the
inflation.
\section{Concluding remarks} There is a well known Polish
ex-minister, a professor of physics, who did not notice a deficit
of a billion in the department under his control though it was
noticed by his sister, a provincial teacher. We remember public
guesses concerning the sources of Bagsik's fortune. We hope that
our arguments limit the inclination towards drawing hasty
conclusions from oversimplified models of financial phenomena.

\end{document}